\documentclass[leqno]{siamltex}
\usepackage{amsmath}

\title{Creation and Annihilation Operators for Orthogonal Polynomials of
Continuous and Discrete Variables} 

\author{Miguel Lorente\thanks{Departamento de F\'{\i}sica, Universidad de Oviedo,
33007 Oviedo, Spain}}

\begin{document}
\maketitle
\pagestyle{myheadings}
\thispagestyle{plain}
\markboth{M. LORENTE}{CREATION AND ANNIHILATION OPERATORS...}
\begin{abstract}
We develop general expressions for the raising and lowering operators that belong
to the orthogonal polynomials of hypergeometric type with discrete and continuous variable. We
construct the creation and annihilation operators that correspond to the normalized polynomials
and study their algebraic properties in the case of the Kravchuk/Hermite Meixner/Laguerre
polynomials.
\end{abstract}

\section{Introduction}

In a previous paper [1] we have developed a method to construct raising and lowering
operators for the Kravchuk polynomials of a discrete variable, using the properties  of
Wigner functions, and to calculate the continuous limit to the creation and annihilation
operators for the solutions of the quantum harmonic oscillator.

In this contribution we apply the same method to other orthogonal polynomials of discrete
and continuous variable. We give general formulas for all orthogonal polynomials of
hypergeometric type [2]: difference/differential equations, recurrence relations, raising
and lowering operators.

With the help of standard values we calculate these equations for the normalized functions
of Kravchuk-Wigner and Meixner-Laguerre polynomials, and we construct the corresponding
creation and annihilation operators. 

The motivation of this work is to implement the study of the Sturm-Liouville
problem in continuous case with the discrete one, in particular the connexion
between the eigenfunctions and the creation and annihilation operators [3] and
[4]. This approach is becoming very powerful in the lattice formulation of
field theories, where the physical properties of the model are analyzed in the
lattice before the continuous limit is taken [5]

\section{Basic relations between orthogonal polynomials of continuous and discrete variable}

A polynomial of hipergeometric type of continuous variable satisfies the following
fundamentals equations

\begin{remunerate}
\item  Differential equation:
\setcounter{equation}{0}
\renewcommand{\theequation}{\mbox{C}{\arabic{equation}}}
\begin{equation}
\sigma \left({s}\right)\ {y''}_{n}\left({s}\right)+\tau
\left({s}\right)\ {y'}_{n}\left({s}\right)+{\lambda }_{n}\ {y}_{n}\left({s}\right)=0
\end{equation}

where $\sigma \left({s}\right)$ and $\tau \left({s}\right)$ are polynomials of
at most second and first degree, respectively and
${\lambda }_{n}$ is a constant. This differential equation can be written in the form of an
eigenvalue equation
\[\left({\sigma \left({s}\right)\rho
\left({s}\right){y'}_{n}\left({s}\right)}\right)'+{\lambda }_{n}\rho
\left({s}\right){y}_{n}\left({s}\right)=0\] where $\rho \left({s}\right)$ is the weight
function satisfying $\left({\sigma \left({s}\right)\rho \left({ s}\right)}\right)'=\tau
\left({s}\right)\rho \left({s}\right)$ and ${\lambda }_{n}=-n\left({\tau
'+{\frac{n-1}{2}}\sigma ''}\right)$.

\item  Orthogonality relations:
\[\int_{a}^{b}{y}_{n}\left({s}\right){y}_{m}\left({s}\right)\rho \left({ s}\right)ds={\delta
}_{nm}{d}_{n}^{2}\]
with ${d}_{n}$ some normalization constant,
\item  Three term recurrence relations:
\begin{equation}
s{y}_{n}\left({s}\right)={\alpha }_{n}{y}_{n+1}\left({s}\right)+{\beta
}_{n}{y}_{n}\left({s}\right)+{\gamma }_{n}{y}_{n-1}\left({s}\right)
\end{equation}
where ${\alpha }_{n}$, ${\beta }_{n}$, ${\gamma}_{n}$ are constants.
\item  Raising and lowering operators:
\begin{equation}
\sigma \left({s}\right){y}_{n}^{'}\left({s}\right)={\frac{{\lambda
}_{n}}{n{\tau }_{n}^{'}}}\left[{{\tau
}_{n}\left({s}\right){y}_{n}\left({s}\right)-{\frac{{B}_{n}}{{B}_{n+1}}}
{y}_{n+1}\left({s}\right)}\right]
\end{equation}
where 
\begin{eqnarray*}
& &{\tau }_{n}\left({s}\right)=\tau \left({s}\right)+n\sigma '\left({s}\right) \\
& &{\tau '}_{n}=\tau '+n\sigma ''=-{\frac{{\lambda }_{2n+1}}{2n+1}}
\end{eqnarray*}
We can modify formula (C3) in a more suitable form.

From 
\[{a}_{n}={B}_{n}\prod\limits_{k\ =\ 0}^{n-1} \left({\tau
'+{\frac{1}{2}}\left({n+k-1}\right)\sigma ''}\right)\ \
\ ,\ \ \ {a}_{0}={B}_{0}\]

we can prove the following identity
\[{\alpha }_{n}={\frac{{a}_{n}}{{a}_{n+1}}}={\frac{{B}_{n}}{{B}_{n+1}}}{\frac{\left({\tau
'+{\frac{n-1}{2}}\sigma ''}\right)}{\left({\tau '\rm +{\frac{2n-1}{2}}\sigma ''}\right)\left({\tau
'+n\sigma ''}\right)}}=-{\frac{2n\left({2n+1}\right)}{{\lambda }_{2n}{\lambda
}_{2n+1}}}{\frac{{\lambda }_{n}}{n}}{\frac{{B}_{n}}{{B}_{n+1}}}\]

that after introducing in (C3) we get a simplified version

\setcounter{equation}{2}
\renewcommand{\theequation}{\mbox{C}{\arabic{equation}}}
\begin{equation}
\sigma \left({s}\right){y'}_{n}\left({s}\right)=-{\frac{{\lambda
}_{n}}{n}}{\frac{2n+1}{{\lambda }_{2n+1}}}{\tau
}_{n}\left({s}\right){y}_{n}\left({s}\right)-{\frac{{\lambda }_{2n}}{2n}}{\alpha
}_{n}{y}_{n+1}\left({s}\right)
\end{equation}

Finally using the recurrence relations (C2) we get
\begin{equation}
\sigma \left({s}\right){y'}_{n}\left({s}\right)=\ \left[{-{\frac{{\lambda
}_{n}}{n}}{\frac{2n+1}{{\lambda }_{2n+1}}}{\tau }_{n}\left({s}\right)-{\frac{{\lambda
}_{2n}}{2n}}\left({s-{\beta }_{n}}\right)}\right]{y}_{n}\left({s}\right)+{\frac{{\lambda
}_{2n}}{2n}}{\gamma }_{n}{y}_{n-1}\left({s}\right)
\end{equation}
\end{remunerate}

Formulas (C3) and (C4) can be used to calculate solutions of the differential equations. In
fact, we put $n=0$ in (C4) and we get a differential equation to calculate $y_0(s)$.

Taking this value in (C3) we obtain by iteration all the polynomials satisfying (C1).

We can implement these formulas in the discrete case. A polynomial of hypergeometric type of
discrete variable satisfies the following fundamental equations.

\begin{remunerate}

\item Difference equation:
\setcounter{equation}{0}
\renewcommand{\theequation}{\mbox{D}{\arabic{equation}}}
\begin{equation}
\sigma \left({x}\right)\Delta \nabla {y}_{n}\left({x}\right)+\tau
\left({x}\right)\Delta {y}_{\rm n}\left({x}\right)+{\lambda }_{n}{y}_{n}\left({x}\right)=0
\end{equation}

where $\sigma \left({x}\right)$ and $\tau \left({x}\right)$ are polynomial of at most second
and first degree, respetively, and the forward (backward) difference operators are:
\[\Delta f\left({x}\right)=f\left({x+1}\right)-f\left({s}\right)\ \ \ ,\ \ \ \nabla 
f\left({x}\right)=f\left({x}\right)-f\left({x-1}\right)\]

This difference equation can be written in the form of an eigenvalue equation
\[\Delta \left[{\sigma \left({x}\right)\rho \left({x}\right)\nabla
{y}_{n}\left({x}\right)}\right]+\lambda \rho \left({x}\right){y}_{n}\left({x}\right)=0\]

where $\rho \left({x}\right)$ is the weight function satisfying
\[\Delta \left[{\sigma \left({x}\right)\rho \left({x}\right)}\right]=\tau
\left({x}\right)\rho \left({\rm x}\right)\]

and
\[{\lambda }_{n}=-n\Delta \tau \left({x}\right)-{\frac{n\left({n-1}\right)}{2}}{\Delta
}^{2}\sigma \left({\rm x}\right)=-n\left({\tau '+{\frac{n-1}{2}}\sigma ''}\right)\]

is the eigenvalue corresponding to the function ${y}_{n}\left({x}\right)$.

\item Orthogonality relations:

The polynomial ${P}_{n}\left({x}\right)$ of hypergeometric
type satisfy the following orthogonal relations:
\[\sum\limits_{2\varepsilon \ =\ a}^{b-1} {P}_{n}\left({x}\right){P}_{m}\left({x}\right)\rho
\left({x}\right)={d}_{n}^{2}{\delta }_{mn}\]

where ${\delta }_{mn}$ is the Kronecker symbol and $d_n$ is some normalization constant.

\item  Three term recurrence relation:
\begin{equation}
x{P}_{n}\left({x}\right)={\alpha
}_{n}{P}_{n+1}\left({x}\right)+{\beta }_{n}{P}_{n}\left({x}\right)+{\gamma
}_{n}{P}_{n-1}\left({x}\right)
\end{equation}

with ${\alpha }_{n}$, ${\beta }_{n}$, ${\gamma}_{n}$ some constants.

\item  Raising and lowering operators:
\begin{equation}
\sigma \left({x}\right)\nabla {P}_{n}\left({x}\right)={\frac{{\lambda
}_{n}}{n{\tau '}_{n}}}\left[{{\tau
}_{n}\left({s}\right){P}_{n}\left({x}\right)-{\frac{{B}_{n}}{{B}_{n+1}}}{P}_{n+1}\left({x}
\right)}\right]
\end{equation}
where 
\begin{eqnarray*}
{\tau }_{n}\left({s}\right) &=&\tau \left({x+n}\right)+\sigma
\left({x+n}\right)-\sigma\left({x}\right) \\
\Delta {\tau }_{n}\left({s}\right)&=&\Delta \tau \left({x}\right)+n{\Delta }^{2}\sigma
\left({x}\right) 
\end{eqnarray*}
or \hspace{35mm}${\tau '}_{n}\;=\quad \tau '+n\sigma
''\left({x}\right)=-{\displaystyle\frac{{\lambda }_{2n+1}}{2n+1}}$

\medskip because $\sigma \left({x}\right)$ and $\tau \left({x}\right)$ are functions of at most
second and first degree, respectively.

We can modify formula (D3) as we did in the continuous case with the help of the identities
\begin{eqnarray*}
{a}_{n}&=&{B}_{n}\prod\limits_{k\ =\ 0}^{n-1} \left({\tau
'+{\frac{1}{2}}\left({n+k-1}\right)\sigma ''}\right), {a}_{0}={B}_{0} \\
{\alpha }_{n}&=&{\frac{{a}_{n}}{{a}_{n+1}}}=-{\frac{2n}{{\lambda
}_{2n}}}{\frac{\left({2n+1}\right)}{{\lambda }_{2n+1}}}{\frac{n}{{\lambda
}_{n}}}{\frac{{B}_{n}}{{B}_{n+1}}}
\end{eqnarray*}

Introducing these identities in (D3) we get a more simplified version:
\setcounter{equation}{2}
\renewcommand{\theequation}{\mbox{D}{\arabic{equation}}}
\begin{equation}
\sigma \left({x}\right)\nabla {P}_{n}\left({x}\right)=-{\frac{{\lambda
}_{n}}{n}}{\frac{\left({2n+1}\right)}{{\lambda }_{2n+1}}}{\tau
}_{n}\left({x}\right){P}_{n}\left({x}\right)-{\frac{{\lambda }_{2n}}{2n}}{\alpha
}_{n}{P}_{n+1}\left({x}\right)
\end{equation}

This expression defines the raising operator ${P}_{n+1}\left({x}\right)$ in terms of the
backward difference operator.

From this expression we can derive another lowering operator in terms of the
forward difference operator. We substitute the operator $\nabla$ in (D3) by its
equivalent difference operator $\nabla \rm =\Delta \rm -\Delta \nabla$ and then
we introduce the difference equation (D1) and the recurrence relation (D2)
obtaining {\setlength{\arraycolsep}{1mm}
\begin{eqnarray}
\qquad\left({\sigma \left({x}\right)+\tau \left({x}\right)}\right)\Delta
{P}_{n}\left({x}\right)\rm &=&\left[{-{\frac{{\lambda }_{n}}{n}}{\frac{2n+1}{{\lambda
}_{2n+1}}}{\tau }_{n}\left({x}\right)-{\lambda }_{n}-{\frac{{\lambda
}_{2n}}{2n}}x+{\frac{{\lambda }_{2n}}{2n}}{\beta
}_{n}}\right]{P}_{n}\left({x}\right)+\\
\nonumber {\smallskip} &+& {\frac{\lambda_{2n}}{2n}}{\gamma }_{n}{P}_{n-1}
\end{eqnarray}}
\end{remunerate}

The advantage of expressions (D3) and (D4) is that all the coefficients are tabulated.

As in the continuous case from (D4), putting $n=0$ we get ${P}_{0}\left({x}\right)$, and
inserting this value in (D3) we obtain the solutions of (D1).

From the orthogonal polynomials of hypergeometric type we can construct the corresponding
normalized functions (up to a phase factor)
\[{\psi }_{n}\left({x}\right)={d}_{n}^{-1}\sqrt {\rho
\left({x}\right)}{P}_{n}\left({x}\right)\]
that satisfy equivalent relations and we denote them (ND1),  (ND2),  (ND3), 
(ND4) in the discrete case and (NC1), (NC2), (NC3), (NC4) in the continuous
case. For instance, (NC1) becomes:
\[\sigma (s) {\psi }''_{n}(s)+\sigma'(s){\psi }'_{n}(s)-\rho
(s)^{-\frac{1}{2}}\left( \sigma (s) \left( \rho
(s)^{\frac{1}{2}}\right)' \right)' {\psi }_{n}(s)+\lambda_n{\psi }_{n}(s)=0\]
which corresponds to a self-adjoint operator of Sturm-Liouville type.
\section{The Hermite and Kravchuk polynomials}

We apply the results of section 2 to some orthogonal polynomials of continuous and discrete
variable and to the corresponding normalized functions.

For the Hermite polynomials ${H}_{n}\left({s}\right)$ we have
\setcounter{equation}{0}
\renewcommand{\theequation}{\mbox{C}{\arabic{equation}}}
\begin{eqnarray}
&&{H''}_{n}\left({s}\right)-2s{H'}_{n}\left({s}\right)+2nH\left({s}\right)=0 \\
&&{sH}_{n}\left({s}\right)={\frac{1}{2}}{H}_{n+1}\left({s}\right)+n{H}_{n-1}\left({s}\right)\\
&&{H}_{n+1}\left({s}\right)=2s{H}_{n}\left({s}\right)-{H'}_{n}\left({s}\right) \\
&&{H}_{n-1}\left({s}\right)={\frac{1}{2n}}{H'}_{n}\left({s}\right)
\end{eqnarray}
Introducing the orthonormalized functions 
\[{\psi }_{n}\left({s}\right)={\left({{2}^{n}n!\sqrt {\pi
}}\right)}^{-1/2}{e}^{-{s}^{2}/2}{H}_{n}\left({s}\right)\]
we get
\setcounter{equation}{0}
\renewcommand{\theequation}{\mbox{NC}{\arabic{equation}}}
\begin{eqnarray}
&&{\psi ''}_{n}\left({s}\right)+\left({2n+1-{s}^{2}}\right){\psi }_{n}\left({s}\right)=0 \\
&&{2s\psi }_{n}\left({s}\right)=\sqrt {2\left({n+1}\right)}{\psi
}_{n+1}\left({s}\right)+\sqrt {2n}{\psi }_{n-1}\left({s}\right) \\
&&\sqrt {n+1}{\psi }_{n+1}\left({s}\right)={\frac{1}{\sqrt
{2}}}\left({s-{\frac{d}{ds}}}\right){\psi }_{n}\left({s}\right) \\
&& \sqrt {n}{\psi }_{n-1}\left({s}\right)={\frac{1}{\sqrt
{2}}}\left({s+{\frac{d}{ds}}}\right){\psi }_{n}\left({s}\right)
\end{eqnarray}

(NC1) describes the quantum harmonic oscillator, (NC3) and (NC4) are the realization of the
familiar creation and annihilation operator
\[{a}^{+}\equiv {\frac{1}{\sqrt {2}}}\left({s-{\frac{d}{ds}}}\right)\ \ \ ,\ \ \ a\equiv
{\frac{1}{\sqrt {2}}}\left({s+{\frac{d}{ds}}}\right)\]

From (NC4) with $n=0$ we obtain ${\psi }_{0}\left({s}\right)$ and inserting this value in
(NC3) we obtain by iteration the solutions of the harmonic oscillator
\[{\psi }_{n}\left({s}\right)={\frac{1}{\sqrt {n!}}}{\left({{a}^{+}}\right)}^{n}{\psi
}_{0}\left({s}\right)\]
In the discrete case we take the Kravchuk polynomials
${k}_{n}^{\left({p}\right)}\left({x,N}\right)$ with $x=0,1,2, \cdots \\ {N-1}$
\begin{eqnarray*}
&&\sigma \left({x}\right)=x \\
&&\tau \left({x}\right) =\left({Np-x}\right)/q \qquad {\tau
}_{n}\left({x}\right)=\left({Np-x-n}\right)/q\;\;+n \\
&&{\lambda }_{n}=n/q \\
&&{\alpha }_{n}=n+1/q \qquad {\beta }_{n}=n+p\left({N-2n}\right), \ \ {\gamma
}_{n}=pq\left({N-n+1}\right)
\end{eqnarray*}

Inserting these values in the fundamental formulas we get

\medskip\noindent (D1)
$p\left({N-x}\right){k}_{n}\left({x+1}\right)+\left[{p\left({n+x-N}\right)+q\left({n-x}\right)}
\right]{k}_{n}\left({x}\right)+qx{k}_{n}\left({x-1}\right)=0$

\medskip\noindent (D2)
$
x{k}_{n}\left({x}\right)=\left({n+1}\right){k}_{n+1}\left({x}\right)+\left[{n+p\left({N-2n}\right)}\right]{k}_{n}\left({x}\right)
+pq\left({N-n+1}\right){k}_{n-1}\left({x}\right)=0$

\medskip\noindent (D3)
$\left({n+1}\right){k}_{n+1}\left({x}\right)=p\left({x+n-N}\right){k}_{n}\left({x}\right)+qx{k}_{n}
\left({x-1}\right) $

\medskip\noindent (D4)
$q\left({N-n+1}\right){k}_{n-1}\left({x}\right)=\left({x+n-N}\right){k}_{n}\left({x}\right)+
\left({N-x}\right){k}_{n}\left({x+1}\right)$

\medskip

For the normalized functions we take the Wigner functions, that appear in the representation
of the rotation group, ${d}_{mm'}^{j}\left({\beta }\right)$
\[{\left({-1}\right)}^{m-m'}{d}_{mm'}^{j}\left({\beta }\right)={d}_{n}^{-1}\sqrt {\rho
\left({x}\right)}{k}_{n}^{(p)}\left({x,N}\right)\]
with \qquad $N=2j$ \qquad $m=j-n$ \qquad $m'=j-x$ \qquad $p ={\sin}^{2}{\displaystyle\frac{\beta
}{2}}\ ,\ q ={\cos}^{2}{\displaystyle\frac{\beta}{2}}$

After substitution we get
\setcounter{equation}{0}
\renewcommand{\theequation}{\mbox{ND}{\arabic{equation}}}
\begin{eqnarray}
\lefteqn {\sqrt{pq\left({N-x}\right)\left({x+1}\right)}{d}_{j-n,\
j-x-1}^{j}\left({\beta}\right) +}\\
\nonumber &+& \left({p\left({N-x-n}\right) + q\left({x-n}\right)}\right){d}_{j-n,\
j-x}^{j}\left({\beta }\right) +\\
\nonumber  &+& \sqrt {pqx\left({N-x+1}\right)}{d}_{j-n,\ j-x+1}^{j}\left({\beta
}\right)=0 \\ [1mm]
\lefteqn{\left[ -p(N-x-n)-q(n-x)\right]{d}_{j-n,\ j-x}^{j}+} \\
\nonumber &+&\sqrt{pq\left({n+1}\right)\left({N-n}\right)}{d}_{j-n-1,\ j-x}^{j}\left({\beta
}\right) +\\
\nonumber &+&\sqrt {pqn\left({N-n+1}\right)}{d}_{j-n+1,\ j-x}^{j}\left({\beta }\right)=0 \\ [1mm]
\lefteqn{\sqrt {pq\left({n+1}\right)\left({N-n}\right)}{d}_{j-n-1,\ j-x}^{j}\left({\beta
}\right) = } \\
\nonumber &=&p\left({N-x-n}\right){d}_{j-n,\ j-x}^{j}\left({\beta }\right)+\sqrt
{pqx\left({N-x+1}\right)}{d}_{j-n,\ j-x+1}^{j} \\ [1mm]
\lefteqn{\sqrt {pqn\left({N-n+1}\right)}{d}_{j-n+1,\ j-x}^{j}\left({\beta
}\right)=} \\
\nonumber &=& p\left({N-x-n}\right){d}_{j-n,\ j-x}^{j}\left({\beta
}\right)+\sqrt {pq\left({x+1}\right)\left({N-x}\right)}{d}_{j-n,\
j-x-1}^{j}\left({\beta }\right)
\end{eqnarray}

The last four equations can be written down in terms of the new parameters $j=N/2\ ,\ \ \
m=j-n\ ,\ \ \ m'=j-x\ ,\ \ \ p={\sin}^{2}\beta / 2\ ,\ \ \ q={\cos}^{2}\beta /2\ ,\ \ \
\sqrt {pq}={\frac{1}{2}}\sin\ \beta $
\setcounter{equation}{0}
\renewcommand{\theequation}{\mbox{ND}{\arabic{equation}}}
\begin{eqnarray}
\lefteqn{\sqrt {\left({j+m'}\right)\left({j-m'+1}\right)}{d}_{m,m'-1}^{j}\left({\beta
}\right)+}\\
\nonumber &+&{\frac{2}{\sin\ \beta }}\left[{m-m'\cos\ \beta }\right]{d}_{m,m'}^{j}\left({\beta
}\right)+ \\
\nonumber &+&\sqrt {\left({j-m'}\right)\left({j+m'+1}\right)}{d}_{m,m'+1}^{j}\left({\beta
}\right)=0 \\ [1mm]
\lefteqn{\sqrt {\left({j+m}\right)\left({j-m+1}\right)}{d}_{m-1,m'}^{j}\left({\beta
}\right) -} \\
\nonumber &-&{\frac{2}{\sin\ \beta }}\left[{m'-m\,\cos\ \beta }\right]{d}_{m,m'}^{j}\left({\beta
}\right)+ \\
\nonumber &+&\sqrt {\left({j-m}\right)\left({j+m+1}\right)}{d}_{m+1,m}^{j}\left({\beta
}\right)=0 \\ [1mm]
\lefteqn{{\frac{1}{2}}\sin \beta \sqrt
{\left({j+m}\right)\left({j-m+1}\right)}{d}_{m-1,m'}^{j}\left({\beta }\right) =} \\
\nonumber &=&{\sin}^{2}{\frac{\beta }{2}}\left({m+m'}\right){d}_{m,m'}^{j}\left({\beta
}\right)+ \\
\nonumber &+&{\frac{1}{2}}\sin\ \beta \sqrt {\left({
j-m'}\right)\left({j+m'+1}\right)}{d}_{m,m'+1}^{j}\left({\beta }\right) \\ [1mm]
\lefteqn{{\frac{1}{2}}\sin\ \beta \sqrt {\left({ j-m}\right)\left({ j+m+1}\right)}{ d}_{ m+1,m'}^{
j}\left({\beta }\right) =} \\
\nonumber &=&{\sin}^{2}{\frac{\beta }{
2}}\left({m+m'}\right){d}_{m,m'}^{j}\left({\beta }\right)+ \\
\nonumber &+&{\frac{1}{2}}\sin\ \beta \sqrt
{\left({ j+m'}\right)\left({ j-m' +1}\right)}{ d}_{ m,m' -1}^{ j}\left({\beta }\right)
\end{eqnarray}

Note that (ND1) and (ND2) are equivalent if we interchange $m\leftrightarrow m'$
 and take in account the general property of Wigner functions
\[{d}_{m,m'}^{j}\left({\beta }\right)={\left({-1}\right)}^{m-m'}{d}_{m' ,m}^{j}\left({\beta
}\right)\]

The same property of duality applies to (ND3) and (ND4).

In Reference [1] we have constructed creation and annihilation operators with the help of
(ND3) and (ND4)
\[{A}^{+}\equiv {\frac{1}{\sqrt {2j}}}\ \sqrt {\left({ j+m}\right)\left({ j-m+1}\right)} \ \ \
,\ \ \ {A}^{-}\equiv {\frac{1}{\sqrt {2j}}}\ \sqrt {\left({ j-m}\right)\left({ j+m+1}\right)}\]
that together with ${A}^{0}\equiv {\frac{1}{2j}}m$, when apply to the spherical functions
${Y}_{jm}$ satisfy the $SO(3)$ algebra 
\[\left[{{A}^{+},{A}^{-}}\right]=2{A}^{0}\ \ \ ,\ \ \ \left[{{A}^{\pm },{A}^{0}}\right]=\pm
{ A}^{\pm } .\]

From the connection between the Wigner functions and the solutions of the quantum harmonic
oscillator we have proved [1] the limit relations
\[{\rm\ (ND3)}\ \rightarrow \ {\rm\ (NC3)}\ \ \ ,\ \ \ {\rm\ (ND4)}\ \rightarrow \ {\rm\ (NC4)}\]

Similar results have been obtained by Bijker et al. [6] for the connection between the
$su(2)$ algebra and the one dimensional anharmonic (Morse) oscillator, and by
Atakishiev [3] for the lattice implementation of the quantum harmonic oscillator.

\subsection{The Laguerre and Meixner polynomials}

Using the general formulas of section 2, we get for the Laguerre polynomials
${L}_{n}^{\alpha }\left({s}\right)$ of continuous variable
\setcounter{equation}{0}
\renewcommand{\theequation}{\mbox{C}{\arabic{equation}}}
\begin{eqnarray}
&&{s{L''}_{n}\left({s}\right)+\left({1+\alpha
-s}\right){L'}_{n}\left({s}\right)+n{L}_{n}\left({s}\right)=0} \\
&&{\left({n+1}\right){L}_{n+1}\left({s}\right)+\left({n+s}\right){L}_{n-1}\left({s}\right)=\left({2n+\alpha
+1-s}\right){L}_{n}\left({s}\right)} \\
&&{s{L'}_{n}\left({s}\right)=-\left({1+\alpha
-s}\right){L}_{n}\left({s}\right)+\left({n+1}\right){L}_{n+1}\left({s}\right)}\\
&&{s{L'}_{n}\left({s}\right)=2n{L}_{n}\left({s}\right)-\left({n+\alpha
}\right){L}_{n-1}\left({s}\right)}
\end{eqnarray}

For the normalized Laguerre functions
\[{\psi }_{n}\left({s}\right)=\sqrt {{\frac{n!}{\Gamma \left({
\alpha +n+1}\right)}}}{e}^{-s/2}{s}^{\alpha /\rm 2}{L}_{n}^{\alpha
}\left({s}\right)\] we obtain the following differential equation, recurrence
relations and expressions for the raising and lowering operators:
\setcounter{equation}{0}
\renewcommand{\theequation}{\mbox{NC}{\arabic{equation}}}

\medskip \noindent (NC1) ${{s}^{2}{\psi }''_{n}\left({s}\right)+s{\psi
}'_{n}\left({s}\right)+{\frac{1}{2}}\left[{-{s}^{2}-{\alpha }^{2}+\alpha s+s}\right]{\psi
}_{n}\left({s}\right)+sn{\psi }_{n}\left({s}\right)=0}$

\medskip \noindent (NC2) ${\sqrt {\left({n+1}\right)\left({n+\alpha +1}\right)}{\psi
}_{n+1}\left({s}\right)+\sqrt {n\left({n+\alpha }\right)}{\psi
}_{n-1}\left({s}\right)=\left({2n+\alpha +1-s}\right){\psi }_{n}^{\alpha }\left({s}\right)}$

\medskip \noindent (NC3) ${\sqrt {\left({n+1}\right)\left({n+\alpha +1}\right)}{\psi
}_{n+1}\left({s}\right)={\frac{1}{2}}\left({2n+\alpha +2-s}\right){\psi
}_{n}\left({s}\right)+s{\psi }'_{n}\left({s}\right)}$

\medskip \noindent (NC4) ${\sqrt {n\left({n+\alpha }\right)}{\psi
}_{n-1}\left({s}\right)={\frac{1}{2}}\left({2n+\alpha -s}\right)-s{\psi
}'_{n}\left({s}\right)}$

\medskip
The first equation, divided by $s$, corresponds to the self-adjoint operator of
the Sturm-Liouville problem for the normalized Laguerre functions.

The last two equations can be considered the creation and annihilation operators for
the normalized Laguerre function. In fact, from (NC4) with n=0 we obtain ${\psi
}_{0}^{\alpha }$. And from (NC3) we easily obtain
 \[{\psi }_{n}^{\alpha }\left({s}\right)={\frac{1}{\sqrt {n!{\left({\alpha
+1}\right)}_{n}}}}{\left({{A}^{+}}\right)}^{n}{\psi }_{0}^{\alpha }\]
where 
\begin{eqnarray*}
A^+\psi_n(s)&=&\sqrt{(n+1)(n+\alpha+1)}\psi_{n+1}(s)\\[2mm]
A^-\psi_n(s)&=&\sqrt{n(n+\alpha)}\psi_{n-1}(s)
\end{eqnarray*}
The Laguerre creation operator was given by Szafraniec [8]. His formula is
equivalent to ours if we substitue (NC1) into (NC3).

Now we take the Meixner polynomials ${m}_{n}^{\gamma }\left({x}\right)$ of discrete variable
$x$, and apply the general expressions of section 2:
\setcounter{equation}{0}
\renewcommand{\theequation}{\mbox{D}{\arabic{equation}}}

\medskip \noindent (D1) $\mu \left({x+\gamma
}\right){m}_{n}\left({x+1}\right)+\left({x-1}\right){m}_{n}\left({x-1}\right)+\left[{-\mu
\left({x+n+\gamma }\right)+n-x}\right]{m}_{n}\left({x}\right)=0$

\medskip \noindent (D2) $\mu {m}_{n+1}\left({x}\right)+n\left({n+\gamma
-1}\right){m}_{n-1}\left({x}\right)=\left[{x\left({\mu -1}\right)+n+\mu
\left({n+\gamma }\right)}\right]{m}_{n}\left({x}\right)$

\medskip \noindent (D3) $\mu {m}_{n+1}\left({x}\right)=\mu \left({x+n+\gamma
}\right){m}_{n}\left({x}\right)-x{m}_{n}\left({x-1}\right)$

\medskip \noindent (D4) $n\left({n+\gamma -1}\right){m}_{n-1}\left({x}\right)=\mu
\left({x+n+\gamma }\right){m}_{n}\left({x}\right)-\mu \left({\gamma
+x}\right){m}_{n}\left({x+1}\right)$

\medskip
For the normalized Meixner polynomials
\[{M}_{n}^{(\gamma )}\left({x}\right)=\sqrt {{\frac{{\mu }^{n}{\left({1-\mu
}\right)}^{\gamma }}{n!{\left({\gamma }\right)}_{n}}}}\sqrt {{\frac{{\mu }^{x}\Gamma
\left({\gamma +x}\right)}{\Gamma \left({x+1}\right)\Gamma \left({\gamma
}\right)}}}{m}_{n}^{\gamma }\left({x}\right)\]
we have the following difference, recurrence equations and raising/lowering operators:
\setcounter{equation}{0}
\renewcommand{\theequation}{\mbox{ND}{\arabic{equation}}}
\begin{eqnarray}
\sqrt {\mu \left({\gamma +x}\right)\left({x+1}\right)}{M}_{n}\left({x+1}\right)&+&
\sqrt{\mu x\left({x+\gamma \rm -1}\right)}{M}_{n}\left({x-1}\right)- \\
\nonumber  &-&\left[{\mu\left({x+n+\gamma }\right)-n+x}\right]{M}_{n}\left({x}\right)=0 \\
\sqrt {\mu \left({\gamma +n}\right)\left({n+1}\right)}{M}_{n+1}\left({x}\right)&+&
\sqrt{\mu n\left({n+\gamma \rm -1}\right)}{M}_{n-1}\left({x}\right)- \\
\nonumber &-&\left[{\mu \left({x+n+\gamma}\right)-x+n}\right]{M}_{n}\left({x}\right)=0 \\
\sqrt {\mu \left({\gamma +n}\right)\left({n+1}\right)}{M}_{n+1}\left({x}\right) &=&
\mu\left({x+n+\gamma }\right){M}_{n}\left({x}\right)- \\
\nonumber &-&\sqrt {\mu x\left({x+\gamma-1}\right)}{M}_{n}\left({x-1}\right) \\
\sqrt {\mu n\left({n+\gamma -1}\right)}{M}_{n-1}\left({x}\right) &=&
\mu \left({x+n+\gamma}\right){M}_{n}\left({x}\right)- \\
\nonumber &-&\sqrt {\mu\left({\gamma+ x}\right)\left({x+1}\right)} {M}_{n}\left({x+1}\right)
\end{eqnarray}
As in the case of Wigner functions there exists for the Meixner normalized functions an
operation of duality between ${\rm\ (D1)}\leftrightarrow \rm (D2)$ and ${\rm\
(D3)}\leftrightarrow {\rm\ (D4)}$ after interchanging $x \leftrightarrow n$.

The last two equations can be consider the creation and annihilation operators for the
normalized Meixner functions:
\setcounter{equation}{2}
\renewcommand{\theequation}{\mbox{ND}{\arabic{equation}}}
\begin{eqnarray}
{A}^{+}{M}_{n}\left({x}\right)&\equiv & \sqrt {\mu \left({\gamma
+n}\right)\left({n+1}\right)}{M}_{n+1}\left({x}\right)\\[1mm]
{A}^{-}{M}_{n}\left({x}\right)&\equiv &\sqrt {\mu n\left({n+\gamma
-1}\right)}{M}_{n-1}\left({x}\right)
\end{eqnarray}
From $A^{-}{M}_{0}\left({x}\right)=0$ and (ND4) we obtain
${M}_{0}\left({x}\right)$ and taking this value into (ND3) we get
\[{M}_{n}\left({x}\right)={\frac{1}{\sqrt {n!{\left({\gamma }\right)}_{n}{\mu
}^{n}}}}{\left({{A}^{+}}\right)}^{n}{M}_{0}\left({x}\right)\]
The operators ${A}^{+},A^{-}$ are adjoint conjugate each other and, together
with
\[{A}^{0}=\mu \left({2n+\gamma }\right)\]
they close the $su(1,1)$ algebra
\[\left[{{A}^{+},{A}^{-}}\right]={A}^{0}\ \ \ ,\ \ \ \left[{{A}^{\pm },{A}^{0}}\right]=\pm
2\mu {A}^{\pm }\]
Finally, from the limit relation between the Meixner and Laguerre polynomials
\[{\frac{1}{n!}}{{m}_{n}}^{\left({\alpha
+1,1-h}\right)}\left({{\frac{s}{h}}}\right)\xrightarrow[h \to 0]{}{L}_{n}^{\alpha }\left({s}\right)\]
it can be proved the limit between the corresponding normalized functions
\[{M}_{n}^{\left({\alpha +1,1-h}\right)}\left({{\frac{s}{h}}}\right)\xrightarrow[h \to 0]{}\
{\psi }_{n}^{\alpha }\left({s}\right)\]
and the limit between the recurrence relations
\[{\rm\ (ND2)}\ \xrightarrow[h \to 0]{}\ {\rm\ (NC2)}\]
The Meixner polynomials of discrete variable can be used in the solutions of the three
dimensional harmonic oscillator in conection with the energy eigenvalues [Ref. 7, formula
(5.10)].

The Laguerre-Meixner creation and annihilation operators were presented by F.H.
Szafraniec at Workshop on Orthogonal Polynomials in Mathematical Physics, June
24-26, 1996, Universidad Carlos III de Madrid, Leganes, Spain, (but not
published).

Atakishiyev has given recently [10] the hamiltonian and the creation and
annihilation operators for the Meixner oscillator. In fact, his formula (43)
and (46) are equivalent to ours (ND1), (ND3), (ND4) for the normalized Meixner
functions. They satisfy the same $su(1,1)$ algebra if we identify $A^+\leftrightarrow
K_+$, $A^-\leftrightarrow K_-$, $A^0\leftrightarrow -K_0$

The Laguerre polynomials of continuous variable are used in the solutions of
the hydrogen atom, with different weight function in the orthogonality
relations [Ref. 7, formula (5.10)].

The wave equation for the hydrogen atom can be considered the differential
equation of the Sturm-Liouville problem for the Laguerre functions 
$$\psi_n(s)\equiv \psi_n^{2l+1}(s)=\sqrt{\rho_1(s)}L_{n-l-1}^{2l+1} (s)$$
where
$$\rho_1(s)=\sigma(s)\rho(s)=s^{2l+2}e^{-s}$$
In fact, after substituion of $\psi_n(s)$ in (C1) for the Laguerre polynomials
we get
\[\psi''_n(s)-\left[ \rho_1(s)\right]^{-\frac{1}{2}}\left( \left[
\rho_1(s)\right]^{\frac{1}{2}}\right)'' \psi_n(s)+(n-l-1)s^{-1} \psi_n(s)=0\]
that corresponds to a self-adjoint operator, from which orthogonality relations
for $\psi_n(s)$ with weight function $s^{-1}$ can be obtained.

\vskip 0.5cm\noindent  { \bf Acknowledgments}
\vskip 0.5cm

The autor want to express his gratitude for allowing him to present this contribution in the
1998 International Workshop on Orthogonal Polynomials. This work has been partially
supported by D.G.I.C.Y.T. Proyecto Pb96-0538. We thank also the Referee for
bringing to our attention References [8] and [9].

\end{document}